\documentclass[prl,twocolumn,amsmath,nofootinbib]{revtex4}

\usepackage{url,graphicx,color}

\newcommand{\deltam}{\delta_{\rm m}}
\newcommand{\deltacb}{\delta_{\rm cb}}
\newcommand{\deltanu}{\delta_{\nu}}
\newcommand{\fcb}{f_{\rm cb}}
\newcommand{\fnu}{f_{\nu}}

\newcommand{\bfv}{\mbox{\boldmath$v$}}

\newcommand{\be}{\begin{equation}}
\newcommand{\ee}{\end{equation}}
\newcommand{\ba}{\begin{eqnarray}}
\newcommand{\ea}{\end{eqnarray}}

\newcommand{\simgt}{\lower.5ex\hbox{$\; \buildrel > \over \sim \;$}}
\newcommand{\simlt}{\lower.5ex\hbox{$\; \buildrel < \over \sim \;$}}

\begin{document}

\title{%
The impact of massive neutrinos on nonlinear matter power spectrum
}%
\author{%
 Shun Saito$^1$, Masahiro Takada$^{2,3}$ and Atsushi Taruya$^{3,4}$
}%
\address{%
 $^1$Department of Physics, The University of Tokyo, Tokyo 113-0033, Japan
}%
\address{%
 $^2$Astronomical Institute, Tohoku University, Sendai 980-8578, Japan
}%
\address{%
 $^3$Institute for the Physics and Mathematics of the Universe (IPMU), 
The University of Tokyo, Chiba 277-8582, Japan
}%
\address{%
 $^4$Research Center for the Early Universe (RESCEU), 
 School of Science, The University of Tokyo, Tokyo 113-0033, Japan
}%

\date{\today}

\begin{abstract}
We present the first attempt to analytically 
study the nonlinear matter power spectrum
for a mixed dark matter (cold dark matter plus neutrinos of
total mass $\sim 0.1$eV) 
model based on cosmological perturbation theory. 
The suppression in the power spectrum amplitudes due to 
massive neutrinos is, compared to the linear regime,
enhanced in the weakly nonlinear regime where standard 
linear theory ceases to be accurate.  
We demonstrate that, thanks to this enhanced effect and the gain in the
range of wavenumbers to which the PT prediction is applicable, 
the use of such a  nonlinear model 
may enable a precision of $\sigma(m_{\nu, {\rm tot}})\sim 0.07$eV 
in constraining the total neutrino mass for 
the planned galaxy redshift survey, a factor of 2 improvement 
compared to the linear regime. 
\end{abstract}

\maketitle


{\bf Introduction}: 
The relic neutrinos having finite masses
cause a characteristic suppression in
the growth of structure formation on scales below
the neutrino free-streaming scale \cite{Bond:1980}. 
Exploring this suppression signature 
from large-scale structure probes can be a vital way to
constrain 
 the neutrino masses 
\cite{Huetal:1998}. In fact the stringent constraints, 
$m_{\nu,{\rm tot}}\simlt 0.2$--$0.6{\rm eV}$, have been derived from
the galaxy power spectrum and the Lyman-$\alpha$ forest power spectrum
\cite{Spergel:2006hy,Tegmark:2006az,Seljak06}. 
Planned galaxy surveys such as the Wide-Field Multi-Object 
Spectrograph (WFMOS) survey \cite{WFMOS} 
will further allow a high-precision measurement of 
the power spectra
and therefore continue to improve the 
sensitivity to neutrino masses (e.g., \cite{Takada:2005si}). 

However, most of the previous work on the subject has been based on 
linear perturbation theory for a mixed dark matter (MDM) model (see
\cite{Lesgourgues:2006nd} for a review).  Even at scales as large as
$\sim 100 $Mpc relevant for the neutrino free-streaming scale,
recent studies based on numerical techniques or perturbation theory 
have shown that the impact of nonlinear clustering cannot be ignored 
for high-precision future surveys, while these studies 
focused mainly on the nonlinear effect on the baryon acoustic oscillations
(BAOs) in the power spectrum \cite{Jeong:2006xd,Taruya:2007}.
Yet, the effects of massive neutrinos are ignored in these studies, 
even though the neutrinos with total mass $\simgt 0.06$eV, implied from the
oscillation experiments, 
cause a $\simgt 5\%$ suppression in the power
spectrum amplitude that surpasses the expected measurement accuracy
($\sim1\%$) at each wavenumber band.
Also unclear is how the neutrino suppression degrades 
the ability of BAO experiments 
for constraining the nature of dark energy as the neutrino effect 
appears at very similar scales to BAOs. 

In this {\it Letter}, we develop a new approach to analytically 
study the nonlinear
power spectrum for a MDM model, based on perturbation theory (PT). PT is
a natural extension of the successful linear theory, and is therefore 
expected to give fairly accurate model predictions up to the
weakly nonlinear regime. We will then use the PT approach to study the
impact of massive neutrinos on the nonlinear power spectrum, 
and discuss how the
use of the PT prediction may help constrain the neutrino
masses for 
WFMOS-like surveys.


{\bf Methodology}:
To develop a PT approach for a MDM model, 
we have to deal with the perturbations of multi-fluid components, 
cold dark matter (CDM), baryon and massive neutrinos, 
which are coupled to each other 
via gravity at
redshifts of interest. Hence the expansion parameter of PT 
is not a single quantity in contrast to the case of a CDM
dominated model in which the expansion parameter is 
only the amplitude of the 
CDM perturbations.
The density perturbation field of total matter is defined as
as 
$\deltam \equiv (\delta\rho_{\rm c}+\delta\rho_{\rm b}+\delta\rho_{\nu})
/\bar{\rho}_{\rm m} = \fcb \deltacb + \fnu \deltanu$, 
where the subscripts `m', `c', `b', `$\nu$' and `cb' stand for total 
matter, CDM, baryon, massive neutrinos, and CDM plus 
baryon, respectively, and
the coefficients, $\fcb$ and $\fnu$, are the fractional
contributions to the matter density, $\Omega_{\rm m0}$: 
$\fnu\equiv \Omega_{\nu0}/\Omega_{\rm m0}=m_{\nu, {\rm
tot}}/(94.1\Omega_{\rm m0}h^2~ {\rm eV})$ and $f_{\rm cb}=1-f_\nu$. 
The total matter power spectrum, $P_{\rm m}(k)$, is then given by
%
\be
 P_{\rm m}(k)=  \fcb^{2}P_{\rm cb}(k) + 2\fcb\fnu P^L_{\rm cb,\nu}(k) 
        + \fnu^{2}P^L_{\nu}(k),\label{eq:Pk_exact}
\ee
%
where the power spectra with the superscript `$L$' denote the 
linear-order spectra (see below) and $P_{\rm cb,\nu}^L$ is the cross 
spectrum between $\deltacb$ and $\deltanu$.  

Mixture of the neutrinos 
in total matter 
affects the nonlinear power spectrum as follows. 
The neutrinos would tend to remain in the linear regime rather than going 
into the nonlinear stage together with CDM and baryon, due to the large
free-streaming. 
In addition, the prefactor $f_\nu$ appearing in Eq.~(\ref{eq:Pk_exact}) 
is likely to be small for a realistic model
(e.g. $\fnu\simlt 0.07$ in \cite{Tegmark:2006az}), allowing the
nonlinear neutrino perturbations to be approximately
ignored. In the following we will thus include only the linear-order
neutrino perturbations, 
which can be accurately computed by solving the
linearized Boltzmann equations \cite{Ma:1995}.
The validity of our assumption will be shown
in \cite{Saito:2007b}\footnote{In brief we have approximately
estimated the nonlinear neutrino perturbations by solving the {\em
modified} Boltzmann equations into which the {\em nonlinear}
gravitational potential including the contribution of the nonlinear
$\delta_{\rm cb}$ 
given by Eq.~(\ref{eq:Pk_OneLoop}) is inserted, 
motivated by the fact that the nonlinear gravitational clustering is
mainly driven by the CDM plus baryon perturbations. As a result, 
the neutrino density perturbation is found to be enhanced only by up to
$\sim 10\%$ for $f_\nu\simlt 0.05$ at scales of interest, corresponding
to less than $0.01\%$ error in the nonlinear power spectrum amplitudes
due to the additional small prefactor $f_\nu$ in
Eq.~(\ref{eq:Pk_exact}).}.

Following the standard PT approach \cite{Makino:1992a}, the CDM plus 
baryon component can be treated as a pressure-less and irrotational 
fluid for the scales of interest.  
The fluid equations for mass and momentum conservation and 
the Poisson equation fully describe the dynamics of the 
density perturbation field, $\deltacb$, and the velocity divergence 
field, $\theta_{\rm cb} \equiv \nabla\cdot \bfv_{\rm cb}/(aH)$. The
solutions to this system can be obtained by making a perturbative expansion, 
$\deltacb=\deltacb^{(1)}+\deltacb^{(2)}+\deltacb^{(3)}+\cdots$ and 
$\theta_{\rm cb}=\theta_{\rm cb}^{(1)}+\theta_{\rm cb}^{(2)}
+\theta_{\rm cb}^{(3)}+\cdots$, where the
superscript `$(i)$' denotes the $i$-th order perturbation. 
In our setting, the nonlinear correction to the total matter power spectrum 
$P_{\rm m}(k)$ arises only through $P_{\rm cb}(k)$ in Eq.~(\ref{eq:Pk_exact}). 
The nonlinear $P_{\rm cb}$ including the 
next-to-leading order corrections is expressed as 
\be
P_{\rm cb}(k;z)=P^{L}_{\rm cb}+P^{(13)}_{\rm cb}+P^{(22)}_{\rm
cb}\ , \label{eq:Pk_cb} 
\ee
where the last two terms describe the nonlinear corrections, the so-called 
one-loop corrections, that include contributions up to the third-order 
perturbations.\par 

\begin{figure}[t]
\begin{center}
\includegraphics[width=0.48\textwidth]{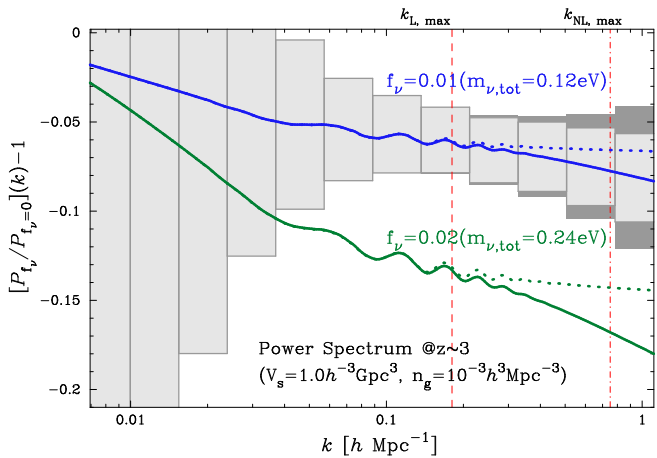}
\end{center}
\vspace*{-2em}
\caption{Fractional difference between the mass power spectra at 
 $z=3$ with 
 and without the massive neutrino contributions, 
where the two cases
 $\fnu=0.01$ and $0.02$ are considered. 
 The solid and dotted curves show the PT and linear theory results, 
 respectively. The two vertical lines indicate a maximum wavenumber 
 limit $k_{\rm max}$ up to which the two models are expected to be valid 
 (see text). The shaded boxes show the expected $1$-$\sigma$ errors on 
 the power spectrum measurement for the $z\sim 3$ WFMOS survey
 and the case of $f_\nu=0.01$. 
}
\label{fig.1}
\end{figure}

The neutrinos affect the spectrum $P_{\rm cb}$ through the
effect on the linear growth rate \cite{Eisenstein:1999jh}. 
At wavenumbers smaller than the neutrino free-streaming scale, 
$k_{\rm fs}(z)\simeq 0.023(m_{\nu}/0.1{\rm eV})
[2/(1+z)]^{1/2}(\Omega_{\rm m0}/0.23)^{1/2}\,h{\rm Mpc^{-1}}$, 
the neutrinos can cluster together with CDM and baryon. 
Conversely, at $k>k_{\rm fs}$, 
the growth rate of CDM perturbations is suppressed due to the weaker 
gravitational force caused by the lack of neutrino 
perturbations. Thus the growth rate, $D_{\rm cb}(z,k)$, has a 
characteristic scale-dependence in a MDM model. 
This fact causes one complication in 
computing the second- and third-order solutions for $\deltacb$ and
$\theta_{\rm cb}$. 
The $k$-dependence of $D_{\rm cb}$ causes mode-couplings 
between the perturbations of different wavenumbers in the nonlinear
regime in addition to the mode-couplings via the transfer function. 
Interestingly, however, 
we have found that, using the analytic fitting formula for 
$D_{\rm cb}$ in \cite{Eisenstein:1999jh}, 
this additional mode-coupling can be safely ignored 
for the expected small value of $f_\nu$ \cite{Saito:2007b}. 
As a result, the nonlinear spectra, $P_{\rm cb}^{(22)}$ and 
$P_{\rm cb}^{(13)}$, 
are written in the form similar to that for a CDM model case 
\cite{Makino:1992a}: 
\begin{eqnarray}
 &&\hspace{-1em}P^{(22)}_{\rm cb}(k;z)=
   \frac{k^{3}}{98(2\pi)^{2}}\int^{\infty}_{0}\!\!dr
   P^{L}_{\rm cb}(kr;z)\nonumber\\
 && \times\int^{1}_{-1}d\mu 
   P^{L}_{\rm cb}(k\sqrt{1+r^{2}-2r\mu};z)
   \frac{(3r+7\mu-10r\mu^{2})^{2}}{(1+r^{2}-2r\mu)^{2}},\nonumber\\
 &&\hspace{-1em}P^{(13)}_{\rm cb}(k;z)=
 \frac{k^{3}P^{L}_{\rm cb}(kr;z)}{252(2\pi)^{2}}
 \int^{\infty}_{0}\!\!dr P^{L}_{\rm cb}(kr;z)
 \left[\frac{12}{r^{2}}-158\right.
\nonumber\\
 &&\left.+100r^{2}-42r^{4}
+\frac{3}{r^{2}}(r^{2}-1)^{3}(7r^{2}+2)
   \ln\left|\frac{1+r}{1-r}\right|\right].\label{eq:Pk_OneLoop}
\end{eqnarray}
Note that $P^{(22)}_{\rm cb}$ and 
$P^{(13)}_{\rm cb}$ are roughly proportional to the square 
of $P^L_{\rm cb}$, which enhances the neutrino effect in the 
nonlinear regime, compared to the linear case, $P^L_{\rm cb}$.


{\bf Results}: Eqs.~(\ref{eq:Pk_exact}) and (\ref{eq:Pk_OneLoop}) show
that the PT prediction for $P_{\rm m}(k)$ at a given redshift can be
computed once the linear spectra, $P_{cb}^L$, $P_{cb,\nu}^L$ and
$P_{\nu}^L$, are specified. We use the CAMB code
\cite{Lewis:1999bs} to compute the input linear spectra for a given MDM 
model\footnote{Our fiducial cosmological parameters 
are $\Omega_{\rm m0}=0.27$ (assuming a flat universe), 
$\Omega_{\rm m0}h^{2}=0.1277,\ 
\Omega_{\rm b0}h^{2}=0.0223,\ n_{s}=1,\ \alpha_{s}=0,\ 
\Delta^{2}_{\cal R}=2.35\times 10^{-9},$ and $w=-1$,
where $n_s$, $\alpha_s$ and $\Delta_{\cal R}^2$ are the 
primordial power spectrum parameters (tilt, running and the 
normalization parameter) and $w$ is the dark energy equation of state.}.
Fig.~\ref{fig.1} shows the fractional difference between the power 
spectra at redshift $z=3$ with and without massive neutrino contributions,
where the two cases $f_\nu=0.01$ 
and 0.02 are considered and 
other parameters are fixed to their fiducial values. 
This plot manifests several interesting points.  First, the 
massive neutrinos induce a characteristic $k$-dependent suppression in 
the spectrum amplitude. For the case of linear theory, 
the suppression becomes nearly independent of $k$ at very small scales, 
$k\gg k_{\rm fs}$, as  roughly given 
by $\Delta P/P \sim -8f_\nu$ \cite{Huetal:1998}. 
In contrast the PT result demonstrates that 
the neutrino suppression 
is {\em enhanced} in 
the nonlinear regime, yielding a new $k$-dependence in the spectrum shape.

Second, comparing the linear theory and PT results explicitly tells
us the limitation of the linear theory: the linear theory is no longer
accurate at $k\simgt 0.2h$Mpc$^{-1}$.  More precisely,
the linear theory result starts to deviate from the PT result at 
$k\simgt k_{\rm L, max}=0.18h$Mpc$^{-1}$ by $\simgt 1\%$ in the amplitude, 
as denoted by the vertical dotted line\footnote{ In Fig.~\ref{fig.1}, 
the deviation of dashed curve (linear) and solid curve (PT) around 
$k_{\rm L, max}$ looks seemingly small due to the fact that, for the
PT result, the denominator 
$P_{\fnu=0}$ in $P_{\fnu}/P_{\fnu=0}$  is also computed from PT.}. 
However PT also breaks down at scales 
greater than a certain maximum wavenumber limit, $k_{\rm NL, max}$, due to 
a stronger mode-coupling arising from the higher-order perturbations 
ignored here. Using $N$-body simulations for a CDM model, 
\cite{Jeong:2006xd} showed that the one-loop PT well matches the simulation
results up to $k_{\rm NL, max}$ given by 
$\Delta^2(k_{\rm NL, max},z)\equiv 
\left.k^3P_m(k,z)/2\pi^2\right|_{k=k_{\rm NL, max}}\simeq 0.4$. 
The vertical dot-dashed line denotes $k_{\rm NL, max}$ derived simply 
assuming this criterion for a MDM model.
Thus, in the case of $z\sim 3$,  
the PT model may allow a factor $4$ gain in $k_{\rm max}$;
observationally, this is roughly equivalent to a factor $64(=4^3)$
gain in independent Fourier modes of the density perturbations probed 
for a fixed survey volume, which in turn improves the 
precision of the power spectrum measurement.

Can a future survey be precise enough to measure the neutrino effect? 
This question is partly answered in Fig.~\ref{fig.1}.  The light-gray 
shaded boxes around the solid curve show the $1$-$\sigma$ measurement 
errors on $P(k)$ at each $k$ bin, expected for the $z\sim 3$ WFMOS survey 
(see below).
The neutrino suppression 
appears to be greater than the errors at 
$k\simgt 0.03h$Mpc$^{-1}$. Another intriguing 
consequence of the nonlinear clustering 
is that the amplified power of $P_m(k)$ reduces the relative 
importance of the shot noise contamination to the measurement errors.
This can be seen by the dark-gray shaded boxes showing the $1$-$\sigma $ 
errors for the linear spectrum.

Finally it would be worth noting that wiggles in the curves
reflect shifts in the BAO peak 
locations caused by the scale-dependent suppression effect
due to neutrinos. The amount of the modulations is smaller than the
measurement errors. 
Hence the uncertainty in neutrino mass
is unlikely to largely degrade the power 
of BAO experiments, at least for an expected small $f_\nu$
\cite{Saito:2007b}.


{\bf Parameter forecasts}: 
To realize the genuine power of future surveys for constraining the 
neutrino masses, we have to carefully take into account parameter 
degeneracies \cite{Takada:2005si}. 
Here we estimate accuracies of the 
neutrino mass determination using the Fisher matrix formalism. \par

The observable we consider is the two-dimensional galaxy power spectrum
given as a function of $k_{\parallel}$ and $k_{\perp}$, the wavenumbers 
parallel and perpendicular to the line-of-sight direction \cite{Seo:2003pu}: 
\be
 P_{s}(k_{{\rm fid}\parallel},k_{{\rm fid}\perp})=
 \frac{D_{A}(z)^{2}_{\rm fid}H(z)}{D_{A}(z)^{2}H(z)_{\rm fid}}
 \left[1+\beta\mu^{2}\right]^{2}b_{1}^{2}P_{\rm m}(k,z)
\ee
where $k=(k_{\perp}^{2}+k_{\parallel}^{2})^{1/2}$ and 
$\mu=k_\parallel/k$. Here,  
$k_{\perp}=[D_{A}(z)_{\rm fid}/D_{A}(z)]k_{{\rm fid}\perp}$ and 
$k_{\parallel}=[H(z)/H(z)_{\rm fid}]k_{{\rm fid}\parallel}$, where 
$D_{A}(z)$ and $H(z)$ are the comoving angular diameter distance and 
Hubble parameter, respectively. The quantities with the subscript `fid' 
denote the quantities estimated assuming a fiducial cosmological model, 
which generally differs from the underlying true model. 
Although the equation above simply assumes
the linear galaxy 
bias $b_1$ and the linear redshift distortion $\beta$,
we will instead 
treat $b_1$ and $\beta$ as free parameters 
in order not to derive too optimistic forecasts. 
This treatment would be adequate for our current purpose, which is 
to estimate how PT allows an improvement in the parameter constraints 
mainly due to the gain in $k_{\rm max}$. 
A more careful analysis will be presented in detail in 
\cite{Saito:2007b}.\par 

\begin{figure}[t]
\begin{center}
\includegraphics[width=0.5\textwidth]{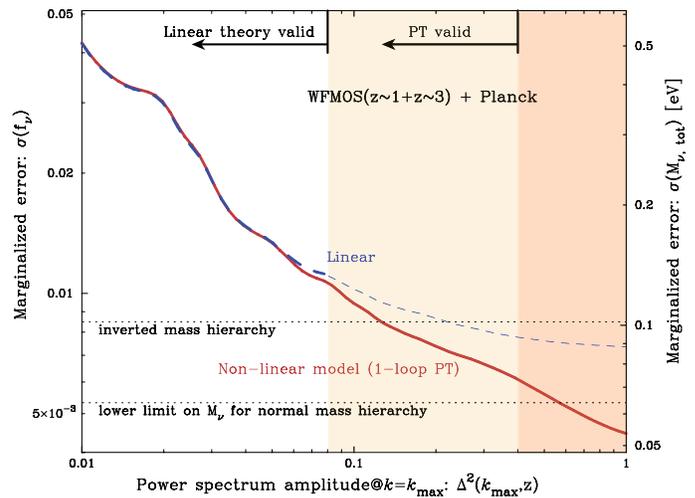}
\end{center}
\vspace*{-2em}
\caption{The marginalized 1-$\sigma$ error on the total neutrino mass as a
 function of the maximum wavenumber $k_{\rm max}$ used in each redshift
 slice (see text), for the WFMOS survey combined with the minimal CMB
 constraints.  The fiducial value of $f_\nu=0.01$ is assumed. 
The solid and dashed curves show the results for the PT
 and linear theory  models, respectively. The light and dark 
shaded regions represent the range of $k$ where the linear theory and
the one-loop PT likely break down due to the stronger nonlinearities. 
}
\label{fig.2}
\end{figure}

Following \cite{Seo:2003pu}, the Fisher matrix for the galaxy power 
spectrum measurement is computed as 
$F_{\alpha\beta}=\int_{-1}^{1}\!d\mu\int_{k_{\rm min}}^{k_{\rm max}}2\pi
k^2dk(\partial P_s/\partial p_\alpha){\rm Cov}^{-1} (\partial
P_s/\partial p_\beta), $
where $p_{\alpha}$ represents a set of parameters and ${\rm Cov}^{-1}$ 
is the inverse of the covariance matrix that depends 
on the power spectrum itself and 
on survey 
parameters, the comoving survey volume and the number density of 
galaxies. To compute $F_{\alpha\beta}$ we need to specify the 
integration range $k_{\rm min}$ and $k_{\rm max}$;
we will throughout 
employ $k_{\rm min}=10^{-4}h$Mpc$^{-1}$ to obtain the fully-convergent
results, and below discuss for the choice 
of $k_{\rm max}$.  Note that, for several redshift slices, we simply add
the Fisher matrices of each slice to obtain the total Fisher matrix. 
The $1$-$\sigma$ error on a certain parameter $p_{\alpha}$ 
marginalized over other parameters is given by 
$\sigma^{2}(p_{\alpha})=({\bf F}^{-1})_{\alpha \alpha}$, where 
${\bf F}^{-1}$ is the inverse of Fisher matrix. 
We employ the WFMOS survey parameters in \cite{Seo:2003pu} 
consisting of two types of redshift surveys: the $z\sim 1$ survey 
covering $0.5\le z\le 1.3$ with 2000 deg$^2$ and the $z\sim 3$ survey 
covering $2.5\le z\le 3.5$ with 300 deg$^2$.  We consider 5 redshift slices.
The choice of free parameters is also important for the Fisher matrix 
formalism: we include a fairly broad range of the model parameters given 
by $p_\alpha= \{\Omega_{\rm m0}, \Omega_{\rm m0}h^2, \Omega_{\rm b0}h^2, 
f_\nu, n_s, \alpha_s, \Delta^2_{\cal R}, w,\beta(z_i),b_1(z_i)\}$. 
We assume three neutrino species that are totally mass 
degenerate and adopt $f_\nu=0.01$ as the fiducial value. 
The fiducial $\beta(z_i)$ 
and $b_1(z_i)$ for the $i$-th redshift slice
are computed following \cite{Seo:2003pu}. In total we 
include 18 free parameters.\par 

Fig.~\ref{fig.2} demonstrates the marginalized $1$-$\sigma$ errors on 
the total neutrino mass as a function of $k_{\rm max}$, where the galaxy 
power spectrum over a range of $k_{\rm min}\le k\le k_{\rm max}$ is included. 
The value of $k_{\rm max}$ for each redshift slice 
is specified by inverting $\Delta^2(k_{\rm max}; z_i)$ for the value 
given in the horizontal axis. 
The errors shown here are for the WFMOS survey combined with 
the CMB information on cosmological parameters except for 
the neutrino masses, $f_{\nu}$, and 
the dark energy parameter,  $w$.
The solid and dashed curves show the 
results for the PT and linear theory, respectively. 
If the linear theory is employed, a reliable accuracy to be obtained
is $\sigma(m_{\nu,{\rm tot}})\simeq 0.13$eV in order not to have a 
biased constraint due to the inaccurate model prediction. 
On the other hand, if the PT prediction is valid up to $\Delta^2(k_{\rm
max})\simeq 0.4$ as discussed in Fig.~\ref{fig.1}, the accuracy of
$\sigma(m_{\nu, {\rm tot}})\simeq 0.072$eV may be attainable, 
a factor of 2 improvement. 

It should be also noted that a wide redshift coverage for 
the planned WFMOS survey is very efficient to break parameter 
degeneracies, especially between the 
neutrino mass and the dark energy parameters 
\cite{Hannestad:2005gj,Takada06}, 
because the dark energy is likely to affect gravitational clustering 
only at low redshifts, $z \simlt 1$.


{\bf Discussion}: It is of great importance to 
carefully study nonlinear structure formation for a most 
realistic model, i.e. a MDM model including $\sim 0.1$eV neutrinos, in
preparation for future galaxy surveys.  While the PT model developed 
in this {\em Letter} gives the first step in this direction, 
another complement to the analytic approach is to use a hybrid 
$N$-body simulation consisting of cold and hot particles, 
which seems feasible with current numerical resources
by extending the pioneering work \cite{Klypin93} for a model 
of $\sim10$eV neutrinos to models
of $\sim 0.1$eV\footnote{After we submitted our paper, 
\cite{Brandbyge08} presented a simulation based study for a MDM model 
where the
similar conclusion, the enhanced neutrino effect in the nonlinear
regime, was found.}.
PT will also play a useful role in 
calibrating/checking the simulations results.

We have demonstrated that the use of PT may enable 
an improvement in the neutrino mass constraint by a factor 2
compared to the case that linear theory is used, 
for the planned WFMOS survey. However our 
study involves several idealizations: most importantly we assumed the linear
galaxy bias and the linear redshift distortion. At least for the large
scales $\sim 100$Mpc, it seems feasible to
develop a self-consistent model to describe galaxy clustering
observables including the non-linear effects on the galaxy bias and
redshift distortions for a MDM model, by using the perturbation theory
approach \cite{Scoccimarro04} and/or the halo model approach and 
by combining with simulations. 
Such a refined model to describe
galaxy clustering observables in the weakly nonlinear regime would
be worth exploring in order to exploit the full potential of the
forthcoming galaxy surveys for constraining or even {\em determining} the
neutrino masses.

{\bf Acknowledgments}: We thank Y.~Suto, O.~Lahav, 
A.~Heavens, and E.~Reese for useful discussion. 
M.T. and A.T. are supported in part by a Grants-in-Aid for
Scientific Research from the JSPS (Nos. 17740129 and 18072001 for MT:
No. 18740132 for AT).


\end{document}